\documentclass[conference]{IEEEtran}
\IEEEoverridecommandlockouts

\usepackage{cite}
\usepackage{amsmath,amssymb,amsfonts}
\usepackage{algorithm}
\usepackage{algorithmic}
\usepackage{graphicx}
\usepackage{textcomp}
\usepackage{xcolor}
\def\BibTeX{{\rm B\kern-.05em{\sc i\kern-.025em b}\kern-.08em
    T\kern-.1667em\lower.7ex\hbox{E}\kern-.125emX}}

 \AtBeginDvi{}
\usepackage{stmaryrd}
\usepackage{bm}
\usepackage{latexsym}
\usepackage{cite}

\newcommand{\jump}[1]{\left\llbracket#1\right\rrbracket}

\usepackage{times}  
\usepackage{helvet}  
\usepackage{courier}  
\usepackage{url}  

\newcommand{\INPUT}{\item[\textbf{Input:}]}
\newcommand{\OUTPUT}{\item[\textbf{Output:}]}
\newcommand{\PROC}{\item[\textbf{Procedure:}]}

\newcommand{\ZZ}{\mathbb{Z}}

\newcommand{\ScaleParam}{q}
\newcommand{\ModulusOfRing}{M}

\begin{document}

\title{MOBIUS: Model-Oblivious Binarized Neural Networks}

\author{\IEEEauthorblockN{Hiromasa Kitai\IEEEauthorrefmark{1}, 
Jason Paul Cruz\IEEEauthorrefmark{1}, 
Naoto Yanai\IEEEauthorrefmark{1}, 
Naohisa Nishida\IEEEauthorrefmark{2}, 
Tatsumi Oba\IEEEauthorrefmark{2}, 
Yuji Unagami\IEEEauthorrefmark{2}, \\ 
Tadanori Teruya\IEEEauthorrefmark{3}, 
Nuttapong Attrapadung\IEEEauthorrefmark{3}, 
Takahiro Matsuda\IEEEauthorrefmark{3}, and 
Goichiro Hanaoka\IEEEauthorrefmark{3}}
\IEEEauthorblockA{\IEEEauthorrefmark{1} 
Osaka University, Japan} 
\IEEEauthorblockA{\IEEEauthorrefmark{2}
Panasonic Corporation, Japan}
\IEEEauthorblockA{\IEEEauthorrefmark{3}
National Institute of Advanced Industrial Science and Technology, Japan}
}

\maketitle

\begin{abstract}
A privacy-preserving framework in which a computational resource provider receives encrypted data from a client and returns prediction results without decrypting the data, i.e., oblivious neural network or encrypted prediction, has been studied in machine learning that provides prediction services. 
In this work, we present \textit{MOBIUS} (Model-Oblivious BInary neUral networkS), a new system that combines \textit{Binarized Neural Networks (BNNs)} and secure computation based on secret sharing as tools for scalable and fast privacy-preserving machine learning. 
BNNs improve computational performance by binarizing values in training to $-1$ and $+1$, while secure computation based on secret sharing provides fast and various computations under encrypted forms via modulo operations with a short bit length. 
However, combining these tools is not trivial because their operations have different algebraic structures and the use of BNNs downgrades prediction accuracy in general. MOBIUS uses improved procedures of BNNs and secure computation that have compatible algebraic structures without downgrading prediction accuracy.
We created an implementation of MOBIUS in C++ using the ABY library (NDSS 2015). We then conducted experiments using the MNIST dataset, and the results show that MOBIUS can return a prediction within 0.76 seconds, which is six times faster than SecureML (IEEE S\&P 2017). 
MOBIUS allows a client to request for encrypted prediction and allows a trainer to obliviously publish an encrypted model to a cloud provided by a computational resource provider, i.e., without revealing the original model itself to the provider.
\end{abstract}

\begin{IEEEkeywords}
Privacy-Preserving Machine Learning, Secure Computation, Neural Network Predictions, Model Obliviousness
\end{IEEEkeywords}

\section{Introduction}

\subsubsection{Background}

Machine learning methods are widely used in various situations, such as healthcare, manufacturing, and financial services. Consequently, privacy has become a serious concern in the use of big data. In general, the following two features are important for practical use of machine learning: \textbf{(1)} make a prediction oblivious without downgrading performance; and \textbf{(2)} guarantee the security of a trained model. 

In the first feature, the application of a privacy-preserving mechanism to a prediction is necessary for guaranteeing the privacy of a client. 
However, a privacy-preserving mechanism may downgrade the throughput of a model and may thus not be used because of poor performance. 
To solve this dilemma, \textbf{a privacy-preserving scheme that does not downgrade performance of a model is necessary}.

In the second feature, different privacy-preserving frameworks have been proposed but \textbf{a framework that hides a model itself, i.e., making the model oblivious, has not been proposed}. 
According to the Recht hypothesis~\cite{Ristenpart17}, \textit{deep neural networks work well because they memorize most of their training data}. 
Several machine learning systems provide prediction as a service in the cloud, and the aforementioned problem strongly affects the trustworthiness of a resource provider who manages a cloud. For example, a resource provider can extract information or even leak a model that he/she receives and manages. Therefore, a trainer who owns a dataset and trains a model has to completely trust a resource provider who provides a prediction service. 
Consequently, a trainer who wants to maintain privacy will hesitate to outsource machine learning services unless he/she completely trusts a resource provider. 
To solve this problem, a model should be encrypted to prevent unauthorized entities, including a resource provider, from accessing the model itself.

\subsubsection{Motivating Example} 

The main goal of this work is to create a system that provides encrypted prediction as well as encryption of a model in such a way that other entities, including a service provider, cannot access the model itself. We call this the \textbf{model-oblivious} problem.

In the model-oblivious problem, there are three entities, namely, a \textit{trainer}, a \textit{resource provider}, and a \textit{client}. 
A trainer trains a model with plaintexts, encrypts the model, and then uploads the encrypted model to a cloud provided by a resource provider. 
When the client utilizes a model, he/she accesses the cloud. 
By encrypting the model, neither the resource provider nor a client can extract information from the model. Similarly, a client can encrypt input data that will be given to the resource provider. 

\begin{figure}[!t]
  \begin{center}
    \includegraphics[width=7cm]{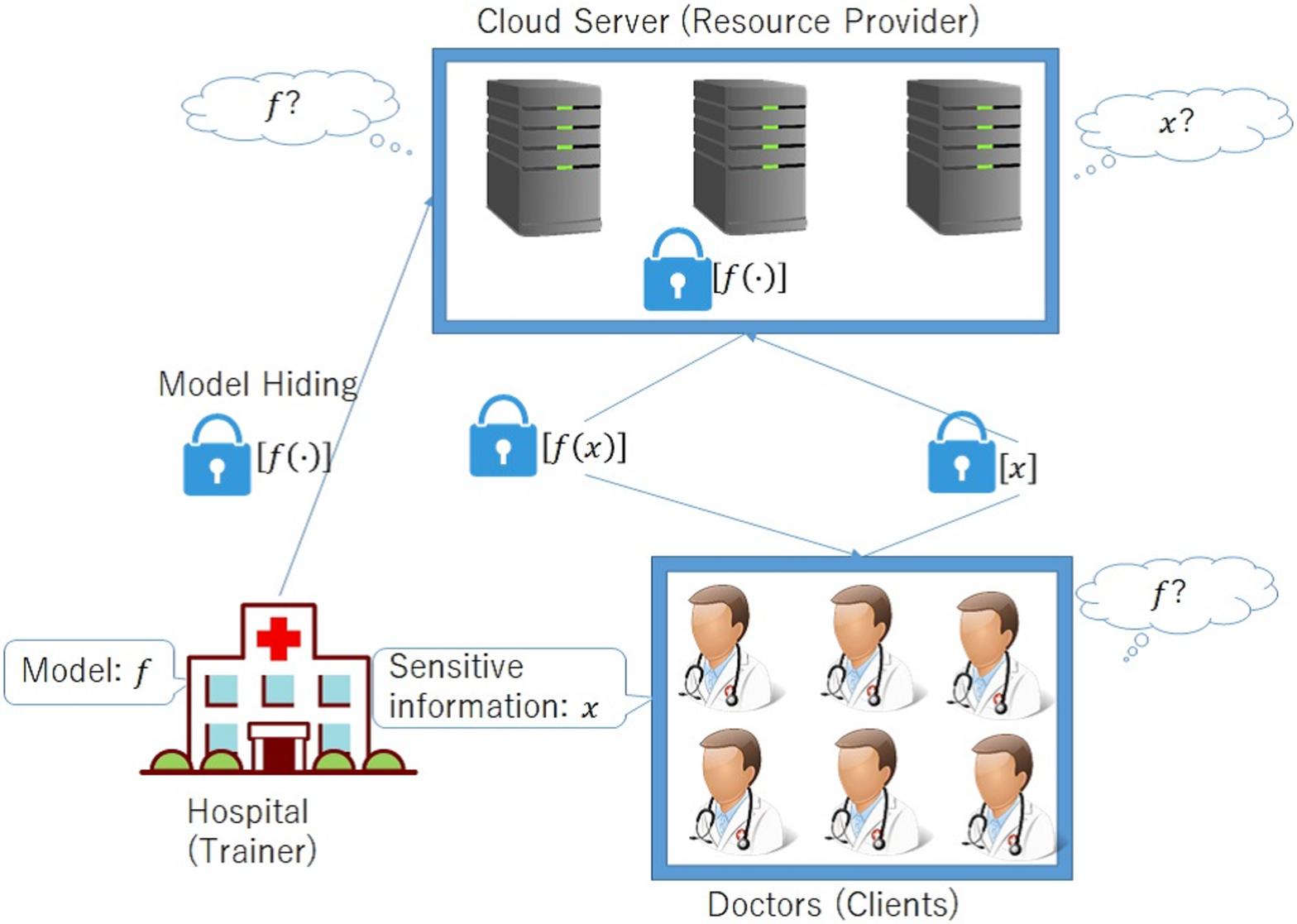}
  \caption{Example Scenario of Model-Oblivious Problem}   
  \label{fig:Example}
  \end{center}
\end{figure}

Figure~\ref{fig:Example} shows an example scenario describing the intuition behind the model-oblivious problem. 
Consider a scenario that includes a hospital, a cloud server, and doctors as the trainer, the resource provider, and clients, respectively. 
The hospital trains a model with datasets it collected, encrypts the model, and then publishes the encrypted model on a cloud server, such as AmazonEC2, to make it publicly available to doctors.
The cloud server can then execute a prediction for an input provided by a doctor by using the encrypted model without decryption. 
With the encrypted model, situations where the cloud server tries to extract information from the model or use the model for other purposes can be prevented. 
In addition, as an equally important measure, the amount of computation required between a client and a resource provider is minimized.

We note that oblivious prediction~\cite{BOP06,OPB07,BPTG15,MiniONN,DeepSecure,EzPC,Chameleon,Gazelle} and encrypted training~\cite{CryptoNets,CGB+17,FHE-DiNN,SecureNN,TAPAS} have not discussed or implied the features of the model-oblivious problem. 
To the best of our knowledge, only SecureML~\cite{SecureML} is the only other system that considers the model-oblivious problem, and our goal is to construct a faster system without sacrificing prediction accuracy.

\subsubsection{Cryptographic Approach} 
In this work, we focus on the use of cryptography for guaranteeing the security of a trained model. 
One of the possible solutions to training while preserving privacy is differential privacy~\cite{Dwork06}, which can prevent a trained model from leaking an individual record by perturbing the records with randomized noise. 
Given this capability, many works on neural networks use differential privacy~\cite{AC+16,CGB+17}. 
There are also works on further applications of differential privacy, e.g., data collection on an untrusted server~\cite{DJW13,DNLA18} or general function release~\cite{AR17}.
However, according to Dowlin et al.~\cite{CryptoNets}, \textit{the notion of differential privacy is not useful in the prediction phase}. 
Moreover, preventing unauthorized entities from accessing a model is outside the scope of differential privacy. 
On the other hand, cryptography can rigorously control authorized access to only users with the correct secret information. 
We therefore construct a system that uses cryptography to encrypt a trained model and prevent unauthorized entities from accessing the trained model.

\subsubsection{Contribution}

In this work, we propose a new system named \textit{Model-Oblivious BInarized neUral networkS (MOBIUS)}, which enables scalable encrypted prediction and encryption of a trained model. 
MOBIUS uses \textit{binarized neural networks (BNNs)}~\cite{BNN} and \textit{secure computation based on secret sharing} as its main tools. 
BNNs are neural networks whose values for weight matrices and activation functions are binarized to $+1$ or $-1$. 
By avoiding the use of real numbers, the computational time of operations with binarized values can be improved. 
Secure computation based on secret sharing distributes input data from a client as shares such that an individual share leaks nothing about the original data, and it can evaluate the data without reconstructing the data via the homomorphism of the shares. 
A bit length of shares can be shortened in comparison with conventional cryptography, and thus the resulting secure computation can perform better than other cryptographic tools, such as fully homomorphic encryption (FHE)~\cite{Gentry09}.

We note that our contribution is \textbf{non-trivial}. We improved the algorithms used in BNNs to make them compatible with the algebraic structures of secure computation. 
The bit-shift method used in the original BNNs~\cite{BNN} downgrades prediction accuracy. 
Batch normalization~\cite{BatchNorm} is used for improving the accuracy but it is based on real numbers and is incompatible with secure computation, which is based on integers. These problems can also potentially downgrade computational performance (See Section \ref{implementation} for details).
To overcome these limitations, we first improve the algorithm of BNNs, particularly the process of batch normalization, to make use of integers and make them compatible with secure computation. 
Both the accuracy and the computational time can be thus improved.

We present the construction of MOBIUS utilizing secure computation based on secret sharing and the improved BNNs and its implementation in C++ using the ABY library~\cite{ABY}. 
We conducted experiments using the MNIST dataset, and the results show that MOBIUS can perform a prediction within 0.76 seconds, which is six times faster than SecureML~\cite{SecureML} even without optimizing our implementation (See Section \ref{result} for details).

\subsubsection{Related Works}

The closest work is SecureML~\cite{SecureML}. 
The main motivation of SecureML was to provide scalable encrypted training, i.e., solving the model-oblivious problem is not their main goal. 
Moreover, encrypted training is the out scope of this work. 
As related works on combining BNNs with cryptography, TAPAS~\cite{TAPAS} and FHE-DiNN~\cite{FHE-DiNN} based on FHE have been concurrently proposed. 
FHE-DiNN utilized discretized neural networks where domains are defined from $-w$ to $+w$, but its experiments were conducted with $-1$ to $+1$ exactly the same as BNNs. 
These works aim to provide fast computation of FHE~\cite{Gentry09,HS15} in BNNs, and they did not discuss the model-oblivious problem.

\section{Preliminaries} \label{Preliminaries}

In this section, we provide backgrounds on neural networks and secure computation to help in understanding our work.

\subsection{Binarized Neural Network} 

Binarized neural networks (BNNs)~\cite{BNN} were proposed to reduce overloads by minimizing data sizes. To do this, values presented in neural networks are binarized to +1 or -1 in order to reduce the required computational resources.

The original work on BNNs~\cite{BNN} described methods to binarize three protocols, namely, \textit{full connection}, \textit{batch normalization}, and \textit{activation}, which are required in standard neural networks. 
Full connection computes matrix multiplications between vectors and weight matrices. 
Batch normalization makes the distribution for nodes uniform in the training phase and contributes to speeding up both training and prediction. 
Activation applies non-linear processing to output vectors, and a sign function is utilized in BNNs.
Among the protocols described above, batch normalization has adopted a bit-shift method to be computed in a binarized form, which is different from well-known batch normalization algorithms~\cite{BatchNorm}, because the operations in well-known batch normalization algorithms require real numbers, consequently creating a bottleneck in the computations.


\newcommand{\SCdomain}{D}
\newcommand{\SCFuncShare}{\mathsf{Share}}
\newcommand{\SCFuncRecon}{\mathsf{Reconst}}

\newcommand{\SCshare}[1]{\llbracket #1 \rrbracket}
\newcommand{\SCprotocol}{\Pi}

\newcommand{\SCclient}{\mathcal{C}}
\newcommand{\SCserver}{\mathcal{S}}

\newcommand{\SCvec}[1]{\boldsymbol{#1}}
\newcommand{\SCenum}[3]{#1_{#2}, \ldots, #1_{#3}}
\newcommand{\SCsetEnum}[3]{\{\SCenum{#1}{#2}{#3}\}}

\newcommand{\SCProtAdd}{\mathsf{ADD}}
\newcommand{\SCProtAddConst}{\mathsf{ADDConst}}
\newcommand{\SCProtMult}{\mathsf{MUL}}
\newcommand{\SCProtMultConst}{\mathsf{MULConst}}
\newcommand{\SCProtCompare}{\mathsf{CMP}}
\newcommand{\SCProtHalf}{\mathsf{Half}}

\subsection{Cryptographic Preliminaries} \label{Crypto. Pre}

In this section, we describe the notations and terminologies used in secure computation based on secret sharing.

\subsubsection{Secret Sharing.}
A $t$-out-of-$n$ secret sharing scheme over a finite domain $\SCdomain$ consists of the following two algorithms:

\begin{itemize}
\item 
$(\SCenum{\SCshare{x}}{1}{n}) \gets \SCFuncShare(x)$: $\SCFuncShare$ takes $x \in \SCdomain$ as input, and outputs $\SCenum{\SCshare{x}}{1}{n} \in \SCdomain$.
            
\item
$x \gets \SCFuncRecon(\SCenum{\SCshare{x}}{1}{t})$: $\SCFuncRecon$ takes $\SCenum{\SCshare{x}}{1}{t} \in \SCdomain$ as input, and outputs $x \in \SCdomain$.
\end{itemize}        
In these algorithms, for $i \in \{1, \ldots, n\}$, $\SCshare{x}_i$ is called the $i$-th share of $x$.
We denote $\SCshare{x} = (\SCenum{\SCshare{x}}{1}{n})$ as their shorthand.
Any less than $t$ shares of $x$ over the $t$-out-of-$n$ secret sharing scheme jointly give no information on $x$, whereas any $\geq t$ shares jointly determine $x$ by using $\SCFuncRecon$.
Several secret sharing schemes that have been proposed typically have finite domains, e.g., a residue class ring $\ZZ_{\ModulusOfRing}$ modulo integer $\ModulusOfRing > 1$ and an $\ell$-length binary string~\cite{Shamir79}.
An $i$-th share of an $\ell$-dimensional vector $\SCvec{v} = (\SCenum{x}{1}{\ell})$ over a domain $\SCdomain$ consists of $i$-th shares of its components and is denoted by $\SCshare{\SCvec{v}}_i := (\SCshare{x_1}_i, \ldots, \SCshare{x_\ell}_i)$.
Analogously, an $i$-th share of a matrix is defined in the same way.
Therefore, a secret sharing scheme over vectors, matrices, and tensors, among others, can be defined.

\subsubsection{Secure Computation based on Secret Sharing.}

We define sub protocols of secure computation that we utilized in our work. 
The following computations are defined over a residue class ring $\ZZ_{\ModulusOfRing} = \{0, \ldots, \ModulusOfRing - 1\}$ modulo integer $\ModulusOfRing$. 
Several efficient implementations of the protocols have been provided~\cite{DBLP:conf/esorics/BogdanovLW08,ABY}.

\begin{itemize}
 \item
$\SCshare{c} \gets \SCProtAdd(\SCshare{a}, \SCshare{b})$: 
$\SCProtAdd$ takes shares $\SCshare{a}$ and $\SCshare{b}$ of $a \in \ZZ_{\ModulusOfRing}$ and $b \in \ZZ_{\ModulusOfRing}$, respectively, as inputs, 
then outputs a share $\SCshare{c}$ of $a + b = c \in \ZZ_{\ModulusOfRing}$.

 \item
$\SCshare{c} \gets \SCProtAddConst(\SCshare{a}, b)$: 
$\SCProtAddConst$ takes share $\SCshare{a}$ of $a \in \ZZ_{\ModulusOfRing}$ and $b \in \ZZ_{\ModulusOfRing}$ as inputs, 
then outputs a share $\SCshare{c}$ of $a + b = c \in \ZZ_{\ModulusOfRing}$.

 \item
$\SCshare{c} \gets \SCProtMult(\SCshare{a}, \SCshare{b})$: 
$\SCProtMult$ takes shares $\SCshare{a}$ and $\SCshare{b}$ of $a \in \ZZ_{\ModulusOfRing}$ and $b \in \ZZ_{\ModulusOfRing}$, respectively, as inputs, 
then outputs a share $\SCshare{c}$ of $a \times b = c \in \ZZ_{\ModulusOfRing}$.

 \item
$\SCshare{c} \gets \SCProtMultConst(\SCshare{a}, b)$: 
$\SCProtMultConst$ takes share $\SCshare{a}$ of $a \in \ZZ_{\ModulusOfRing}$ and $b \in \ZZ_{\ModulusOfRing}$ as inputs, 
then outputs a share $\SCshare{c}$ of $a \times b = c \in \ZZ_{\ModulusOfRing}$.

 \item
$\SCshare{c} \gets \SCProtCompare(\SCshare{a}, \SCshare{b})$: 
$\SCProtCompare$ takes shares $\SCshare{a}$ and $\SCshare{b}$ of $a \in \ZZ_{\ModulusOfRing}$ and $b \in \ZZ_{\ModulusOfRing}$, respectively, as inputs, 
then outputs a share $\SCshare{1}$ if $a < b$ over the integers, $\SCshare{0}$ otherwise.

 \item
$\SCshare{c} \gets \SCProtHalf(\SCshare{a})$: 
$\SCProtHalf$ takes a share $\SCshare{a}$ of $a \in \ZZ_{\ModulusOfRing}$ as input, 
then outputs a share $\SCshare{1}$ if $a \leq \lfloor \ModulusOfRing / 2 \rfloor$ over the integers, $\SCshare{0}$ otherwise.
\end{itemize}

In our implementation, we utilize the ABY library~\cite{ABY}, which is based on a two-party setting (See Section \ref{implementation} for details), and supports secure computation over a residue class ring modulo integer $\ModulusOfRing = 2^m$~($m = 8$, $16$, $32$, or $64$).
Here, $\SCProtHalf$ can be instantiated by the use of $\SCProtCompare$ although it is not originally included in the ABY library. 


\subsection{Security and Network Settings} \label{Security Setting}

In this paper, we focus on the semi-honest adversary.
More precisely, we consider the adversary who follows protocols but curiously learn client's or trainer's data.
As mentioned above, in our proposed protocol, there are three parties: the client, the service provider, and the trainer, and note that there are $n$ servers in the cloud hosted by the service provider.

The trainer locally trains with plaintexts, i.e., non-encrypted training, and constructs a model of a BNN.
Then the trainer computes shares of the model with respect to an underlying $t$-out-of-$n$ secure computation scheme, and then uploads the resulting shares to the servers.
Namely, the adversary cannot learn the model as long as the adversary corrupts less than $t$ servers.

The client computes shares of its query of the prediction on trainer's model with respect to the underlying $t$-out-of-$n$ secure computation scheme,
and then sends the resulting shares to the cloud.
More than $t - 1$ servers jointly compute a protocol of the prediction with input the shares of model and the shares of query, and then output its result.
Namely, the adversary cannot learn client's query as long as the adversary corrupts less than $t$ servers.

However, similar to previous proposals~\cite{SecureML,MiniONN}, we do not aim to hide the size of client's query, the network architecture of trainer's model, and which secure computation protocols are used. 
The authors of MiniONN~\cite{MiniONN} suggested that such information can be protected by adding dummy layers, which can also be integrated with our proposed protocol. 

Finally, we assume the use of secure channel, which can be instantiated by the transport layer security (TLS)~\cite{rfc5246}. 
This setting is the same as that in other literature~\cite{SecureML,MiniONN}.




\section{Our Main Idea}

\subsubsection{Technical Problem}

This work aims to create a system that achieves both the performance and the security of a trained model by using BNNs.
The values in the operations of BNNs are binarized into $+1$ or $-1$ and may seem to be compatible with the algebraic structures of secure computation. 
However, the processes of the original batch normalization~\cite{BatchNorm} that improve the performance of neural networks are linear operations in real numbers, making them incompatible with secure computation in integers.

\subsubsection{Transformation into Integers}

To solve the compatibility problem, we transform the parameters of batch normalization into linear operations in integers by truncating lower digits of the parameters and then multiplying them by a constant. 
We heuristically know that such transformation has small influence on the accuracy because errors can be reset by using non-linear processing in an activation function after the batch normalization. 
In particular, the possibility that the truncation of digits changes the sign of the output of batch normalization (i.e., from positive to negative and vice versa) and influence the activation function is negligible.
The output of the batch normalization in the output layer is identical to that of BNNs, and the maximized value in these output vectors can be finally obtained as a prediction result. 
The possibility that the index of a maximized value is changed is negligible, and thus the truncation of digits does not affect the prediction result.
In actual applications, the sizes of the parameters can be chosen such that the decline in the accuracy in a trained BNN model is minimal.

The method described above solves the incompatibility problem between the algebraic structures of the operations of BNNs and secure computation. 
Moreover, this method achieves a higher accuracy than the bit-shift method in the original BNNs~\cite{BNN} because the standard batch normalization can clip distribution with a higher accuracy. 
Finally, we can construct MOBIUS by combining an efficient and scalable secure computation based on secret sharing and the improved BNNs.

\section{Binarized Neural Networks Compatible with Secure Computation}

In this section, we propose improved BNNs to be used in MOBIUS.
First, we discuss the difference between the proposed BNNs and the original BNNs~\cite{BNN}.
Then, we describe the algorithms used in the proposed BNNs.
Finally, we instantiate an architecture for MNIST, a large database of handwritten digits,
as a concrete example of the proposed BNNs.

\subsection{Avoiding Shift-Based Batch Normalization}
\label{ProposedBatchNorm}

As described in the previous section, our batch normalization uses only integers.
Let $\gamma_{(i)}, \beta_{(i)}, \mu_{(i)},$ and $\sigma_{(i)}$ be learned parameters and $\epsilon$ a small positive value.
The result of ordinary batch normalization can be obtained with the following equation:
\begin{eqnarray} \label{equation 1}
\hat{x}_{(i)} = \gamma_{(i)} \frac{x_{(i)} - \mu_{(i)}}{\sqrt{\sigma_{(i)}^{2} + \epsilon}} + \beta_{(i)} . 
\end{eqnarray}
By replacing coefficients, Equation (\ref{equation 1}) can be transformed as 
$\hat{x}_{(i)} = s_{(i)}x_{(i)} + t_{(i)}$ where
\begin{eqnarray} \label{computation s and t}
s_{(i)} = \frac{\gamma_{(i)}}{\sqrt{\sigma_{(i)}^{2} + \epsilon}}, \ 
t_{(i)} = \beta_{(i)} - \frac{\gamma_{(i)} \mu_{(i)}}{\sqrt{\sigma_{(i)}^{2} + \epsilon}} .
\end{eqnarray}
By substituting $s'_{(i)}, t'_{(i)}$ for integers $s_{(i)}, t_{(i)}$ using an appropriate integer $\ScaleParam$, called scale parameter,
we obtain an alternative integer $\hat{x}'_{(i)}$ for $\hat{x}_{(i)}$ as follows: 
\begin{eqnarray}
\hat{x}'_{(i)} = s'_{(i)} x_{(i)} + t'_{(i)} \ \left(s'_{(i)} = \lfloor qs_{(i)} \rfloor, t'_{(i)} = \lfloor qt_{(i)} \rfloor\right)
\end{eqnarray}
Although the value of $\ScaleParam$ can be determined layerwise or even nodewise, the same $\ScaleParam$ is used in every node for brevity in this paper. 
As the value $\ScaleParam$ increases, the deterioration of BNN prediction accuracy decreases. 
However, the increase in $\ScaleParam$ causes the increase of a bit length of a modulo $\ModulusOfRing$. 
However, the value of the bit length of modulo $\ModulusOfRing$ increases as $\ScaleParam$ increases, consequently increasing memory requirements and calculation costs. 
Therefore, $\ScaleParam$ should be as small as possible to maintain high prediction accuracy. 

\subsection{Improved Binarized Neural Networks}
\label{modifiedBNN}

In this section, we describe the binary full connection, batch normalization, and activation algorithms used in the proposed BNNs.
The binary full connection algorithm is shown in Algorithm~\ref{algbfc}.
This algorithm takes an integer vector $\bm{a}$ and a learned weight matrix $\bm{W}$ as inputs, then outputs
the result of matrix multiplication $\bm{Wa}$.
\begin{algorithm}[htb]
	\caption{BinaryFullConnection}
	\label{algbfc}
	\begin{algorithmic}[1]
{\small
		\INPUT $\bm{a} \in \ZZ^{d_{in} \times 1}$ : input vector\\
				$\bm{W} \in \{-1,1\}^{d_{out} \times d_{in}}$ : weight matrix
		\OUTPUT $\bm{c} \in \ZZ^{d_{out} \times 1}$
		\PROC
		\STATE $\bm{c} \leftarrow \bm{W} \bm{a}$
}
	\end{algorithmic}
\end{algorithm}

The batch normalization algorithm is shown in Algorithm~\ref{algbbn}.
This algorithm takes an integer vector $\bm{c}$, which is usually an output of binary full connection,
and batch normalization parameters $s', t'$ as inputs, then outputs
the result of batch normalization.
Batch normalization parameters $s', t'$ are obtained as described in \ref{ProposedBatchNorm}.

\begin{algorithm}[h]
	\caption{BatchNormalization}
	\label{algbbn}
	\begin{algorithmic}[1]
	{\small 
		\INPUT $ \bm{c} \in \ZZ^{d} $ : input vector\\
				$\bm{s',t'} \in \ZZ^{d}$: batch normalization parameters
		\OUTPUT $\bm{b} \in \ZZ^d$
		\PROC
		\STATE \textbf{for $i=1$ to $d$}, $b_{(i)} \leftarrow s'_{(i)} * c_{(i)} + t'_{(i)}$
}
	\end{algorithmic}
\end{algorithm}

The activation algorithm is shown in Algorithm~\ref{algbact}.
This algorithm takes an integer vector $\bm{b}$ as input,
then outputs a binary vector that represents the signs of each element of the input vector $\bm{b}$.
\begin{algorithm}[h]
	\caption{Activation}
	\label{algbact}
	\begin{algorithmic}[1]
	{\small
		\INPUT $ \bm{b} \in \ZZ^{d} $: input vector
		\OUTPUT $\bm{a} \in \{-1,1\}^{d}$
		\PROC
			\STATE \textbf{for $i=1$ to $d$},
$a_{(i)} \leftarrow \begin{cases}
						-1 & (b_{(i)} < 0) \\
						 1 & (b_{(i)} \geq 0)
					\end{cases}$
}
	\end{algorithmic}
\end{algorithm}

\subsection{Binarized Neural Networks for MNIST dataset} \label{BNN for MNIST}
In Section \ref{modifiedBNN}, we described the algorithms used in the proposed BNNs.
To use BNNs for learning or predicting data, we need to instantiate a concrete architecture and
determine the entire procedure.
We instantiate an architecture for MNIST dataset image classification (See Section~\ref{Data Sets} for details on the MNIST dataset).

\begin{figure}[!t]
  \begin{center}
    \includegraphics[width=9cm]{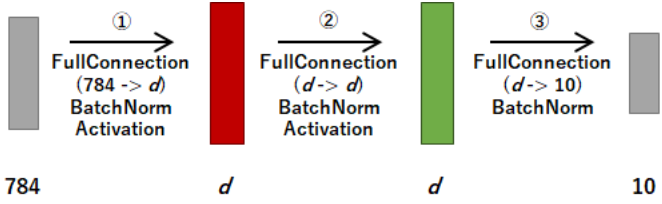}
  \caption{Architecture of the BNNs for MNIST dataset}
  \label{fig:bnn_for_mnist}
  \end{center}
\end{figure}

Consider a typical architecture with an input layer of size 784, two hidden layers of size $d$, and an output layer of size 10, as shown in Figure~\ref{fig:bnn_for_mnist}.
In the hidden layers, the full connection, batch normalization, and activation algorithms
are executed in order. 
In the output layer, only the full connection and batch normalization algorithms are executed.
In this architecture, even though the maximum value index of the output vector is the result of the prediction, we omit this process because the proposed method is designed to return output vectors as the result of secure computation.

\begin{algorithm}[htb]
	\caption{Binarized Neural Network for MNIST}
	\label{algbnn}
	\begin{algorithmic}[1]
	{\small
		\INPUT $ \bm{input} \in \ZZ^{784 \times 1}$ : input vector \\
		$\bm{W_1} \in \{-1, 1\}^{d \times 784}, \bm{s'_{1}} \in \ZZ^{d \times 1}, \bm{t'_1} \in \ZZ^{d \times 1},$\\
		$\bm{W_2} \in \{-1, 1\}^{d \times d}, \bm{s'_2} \in \ZZ^{d \times 1}, \bm{t'_2} \in \ZZ^{d \times 1},$\\
		$\bm{W_3} \in \{-1, 1\}^{10 \times d}, \bm{s'_3} \in \ZZ^{10 \times 1}, \bm{t'_3} \in \ZZ^{10 \times 1}$\\
		\OUTPUT $\bm{output} \in \ZZ^{ 10 \times 1 }$ : output vector
		\PROC
		\STATE $\bm{c_1} \leftarrow \mathsf{FullConnection} \left( \bm{input}, \bm{W_1} \right)$
		\STATE $\bm{b_1} \leftarrow \mathsf{BatchNormalization} \left( \bm{c_1}, \bm{s'_1}, \bm{t'_1}  \right)$
		\STATE $\bm{a_1} \leftarrow \mathsf{Activation} \left( \bm{b_1} \right)$


		\STATE $\bm{c_2} \leftarrow \mathsf{FullConnection} \left( \bm{a_1}, \bm{W_2} \right)$
		\STATE $\bm{b_2} \leftarrow \mathsf{BatchNormalization} \left( \bm{c_2}, \bm{s'_2}, \bm{t'_2}  \right)$
		\STATE $\bm{a_2} \leftarrow \mathsf{Activation} \left( \bm{b_2} \right)$


		\STATE $\bm{c_3} \leftarrow \mathsf{FullConnection} \left( \bm{a_2}, \bm{W_3} \right)$
		\STATE $\bm{output} \leftarrow \mathsf{BatchNormalization} \left( \bm{c_3}, \bm{s'_3}, \bm{t'_3}  \right)$
		}
	\end{algorithmic}
\end{algorithm}
The weight matrices $\bm{W_1}, \bm{W_2},$ and $\bm{W_3}$ used in algorithm~\ref{algbnn} are learned parameters, and
the batch normalization parameters $\bm{s'_j}, \bm{t'_j}$ can be calculated as described in Section~\ref{ProposedBatchNorm}.
In the case of $d = 128, 1000$ ($d$ is the size of hidden layers), 
we confirm experimentally that the deterioration of prediction accuracy towards test data is negligible when a scale parameter $\ScaleParam = 10,000$. 
Therefore, we use $\ScaleParam = 10,000$ in all experiments in this work.

\section{MOBIUS Design}

In this section, we describe the design of MOBIUS. 
We first describe share generation of a trained model in the pre-processing phase, and then show its main algorithms.

MOBIUS is composed of protocols we call \textit{secure full connection}, \textit{secure batch normalization}, and \textit{secure activation}. The main sequences of these protocols are almost the same as those described in the previous section, but we utilize secure computation in the internal processes.

\subsection{Secret Sharing a Model}

We first construct shares of parameters, which are learned in plaintexts, except for that of batch normalization by utilizing secret sharing described in Section~\ref{Crypto. Pre}. 
In this construction, let $\ModulusOfRing$ be a modulo of the secret sharing. 
Moreover, for any $a$, $\jump{ \bm{ a} }$ is a secret share if $a > 0$ and $ \jump{ \bm{\ModulusOfRing  + a} }$ is a secret share if $a < 0$. Hereinafter, we denote $0 \leq a \leq \lfloor \frac{\ModulusOfRing}{2} \rfloor$ as a non-negative integer and $\lfloor \frac{\ModulusOfRing}{2} \rfloor < a < \ModulusOfRing$ as a negative integer.

Learned weight matrices $W_i$ ($i = 1, \cdots ; L -1$) are shared using secret sharing and are stored in each server in a distributed manner. 
Parameters of the batch normalization are computed using the computation in Section~\ref{ProposedBatchNorm}, and its resulting parameters $s_i$, $t_i$ ($i = 1, \cdots,  L-1$) are stored in each sever as shares by utilizing the secret sharing. 
Finally, the size information $(L,  n_0, \cdots, n_L)$ of the shares themselves are not shared, i.e., they are stored as plaintexts.

\subsection{Model-Oblivious Prediction}
\label{sec:proposeprotocol}

The construction of a prediction protocol for MNIST in MOBIUS is shown in Algorithm~\ref{algsecbcnn}. 
The secure full connection, secure batch normalization, and secure activation are denoted by $\mathsf{SecureFC}, \mathsf{SecureBN},$ and $\mathsf{SecureAct}$, respectively. 
Moreover, for any matrix $\bm{X}$, $\bm{X_{i,j}}$ indicates an element of the $i$-th row and $j$-th column and $\bm{X_{i} }$ indicates an element of the $i$-th column. 
\begin{algorithm}[htb]
	\caption{SecureBinaryNN for MNIST}
	\label{algsecbcnn}
	\begin{algorithmic}[1]
	{\small
		\INPUT $\jump{ \bm{input} } \in \ZZ_{\ModulusOfRing}^{784}$: Shares of Input Vectors\\
		\OUTPUT $\jump{\bm{output}} \in \ZZ_{\ModulusOfRing}^{10}$: Prediction Results
		\PROC
		\STATE $\jump{\bm{c_1}} \leftarrow \mathsf{SecureFC} \left( \jump{\bm{input}}, \jump{\bm{W_1}} \right)$
		\STATE $\jump{\bm{b_1}} \leftarrow \mathsf{SecureBN} \left( \jump{\bm{c_1}}, \jump{\bm{s_1}}, \jump{\bm{t_1}}  \right)$
		\STATE $\jump{\bm{a_1}} \leftarrow \mathsf{SecureAct} \left( \jump{\bm{b_1}} \right)$


		\STATE $\jump{\bm{c_2}} \leftarrow \mathsf{SecureFC} \left( \jump{\bm{a_1}}, \jump{\bm{W_2}} \right)$
		\STATE $\jump{\bm{b_2}} \leftarrow \mathsf{SecureBN} \left( \jump{\bm{c_2}}, \jump{\bm{s_2}}, \jump{\bm{t_2}}  \right)$
		\STATE $\jump{\bm{a_2}} \leftarrow \mathsf{SecureAct} \left( \jump{\bm{b_2}} \right)$


		\STATE $\jump{\bm{c_3}} \leftarrow \mathsf{SecureFC} \left( \jump{\bm{a_2}}, \jump{\bm{W_3}} \right)$
		\STATE $\jump{\bm{output}} \leftarrow \mathsf{SecureBN} \left( \jump{\bm{c_3}}, \jump{\bm{s_3}}, \jump{\bm{t_3}}  \right)$
}
	\end{algorithmic}
\end{algorithm}

The secure full connection protocol is described in Algorithm~\ref{algsecfc}. 
The matrix multiplication between shares is computed similarly as in Algorithm~\ref{algbfc}. 
\begin{algorithm}[h]
	\caption{SecureFullConnection}
	\label{algsecfc}
	\begin{algorithmic}[1]
	{\small
		\INPUT $\jump{\bm{input}} \in \ZZ_{\ModulusOfRing}^{d_{in}}$: Shares of Input Vectors\\
				$\jump{\bm{W}} \in \ZZ_{\ModulusOfRing}^{d_{out} \times d_{in}}$: Shares of Weight Matrices\\
		\OUTPUT $\jump{ \bm{output} } \in \ZZ_{\ModulusOfRing}^{d_{out}}$
		\PROC
		\FOR{$i = 0$ to $d_{out}$}
			\FOR{$j = 0$ to $d_{in}$}
				\STATE $\jump{ \bm{X_{i}} } \leftarrow  \SCProtMult(\SCshare{ \bm{W_{i,j}} }, \SCshare{\bm{input_{j}}} ) $
				\STATE $\jump{ \bm{output_{i}} } \leftarrow  \SCProtAdd(\SCshare{ \bm{output_{i}} }, \SCshare{\bm{ X_i}} ) $
			\ENDFOR
		\ENDFOR
		}
	\end{algorithmic}
\end{algorithm}

The secure batch normalization protocol is described in Algorithm~\ref{algsecbn}. 
Although the original batch normalization~\cite{BatchNorm} requires computations of root or division, the secure batch normalization protocol can be performed with only addition and multiplication by performing the computation in Equation~(\ref{computation s and t}) in advance. 
\begin{algorithm}[h]
	\caption{SecureBatchNormalization}
	\label{algsecbn}
	\begin{algorithmic}[1]
	{\small
		\INPUT $\jump{\bm{c}} \in \ZZ_{\ModulusOfRing}^{d}$: Shares of Input Vectors\\
				$\jump{\bm{s}}, \jump{\bm{t}} \in \ZZ_{\ModulusOfRing}^{d}$: Batch Normalization Parameters
		\OUTPUT $\jump{\bm{output}} \in \ZZ_{\ModulusOfRing}^{d}$
		\PROC
			\FOR{$j = 0$ to $d$}

				\STATE $\jump{ \bm{X_{j}} } \leftarrow  \SCProtMult(\SCshare{ \bm{c_j} }, \SCshare{\bm{s_{j}}} ) $
				\STATE $\jump{ \bm{output_{i}} } \leftarrow  \SCProtAdd(\SCshare{ \bm{X_{j}} }, \SCshare{\bm{ t_j}} ) $
			\ENDFOR
			}
	\end{algorithmic}
\end{algorithm}

The secure activation protocol is described in Algorithm~\ref{algsecact}. 
This algorithm outputs $+1$ if the input is greater than or equal to zero or $-1$ otherwise. 
As described above, a non-negative integer is represented by $\{0, \ldots, \lfloor \frac{\ModulusOfRing}{2} \rfloor\}$ and a negative integer is represented by $\{\lfloor \frac{\ModulusOfRing}{2} \rfloor + 1, \ldots, \ModulusOfRing - 1\}$. 
Therefore, the algorithm is performed by a comparison operation with $\lfloor \frac{\ModulusOfRing}{2} \rfloor + 1$ in secure computation.
\begin{algorithm}[h]
	\caption{SecureActivation}
	\label{algsecact}
	\begin{algorithmic}[1]
	{\small
		\INPUT $\jump{\bm{b}} \in \ZZ_{\ModulusOfRing}^{d}$: Shares of Input Vectors\\
		\OUTPUT $\jump{ \bm{a} } \in \ZZ_{\ModulusOfRing}^{d}$
			\FOR{$j = 0$ to $d$}
					\STATE $\SCshare{X_j} \gets \SCProtHalf(\SCshare{b_j})$
					\STATE $\SCshare{Y_j} \gets \SCProtMultConst(\SCshare{X_j}, 2)$
					\STATE $\jump{\bm{a_{j}}} \leftarrow \SCProtAddConst(\SCshare{Y_j}, -1)$
			\ENDFOR
			}
	\end{algorithmic}
\end{algorithm}

\begin{table*}[t!]
\centering
  \begin{tabular}{l|c|c|c}
     & Model Obliviousness & Accuracy [\%] & Time [sec]\\ \hline \hline
    SecureML~\cite{SecureML} & \checkmark &  93.1 & 4.88 \\ \hline
    MiniONN~\cite{MiniONN} & & 97.6  & 1.04 \\ \hline
    TAPAS~\cite{TAPAS} &  & 97.3 & 147 \\ \hline
    FHE-DiNN~\cite{FHE-DiNN} & & 96.3 & 1.64 \\ \hline
    MOBIUS & \checkmark &  95.9 & 0.76 \\
  \end{tabular}
\caption{Performance of MOBIUS with MNIST dataset in comparison with related works.} 
The second column refers to the capability to make a model oblivious. 
The third column refers to the evaluation of the model as output at the prediction phase. The final column refers to total elapsed time of execution in the real time, i.e., including both off-line computation and on-line computation. 
Although several source codes of the other systems were obtained, we could not execute them and the values in the Table were obtained from their papers.
\label{Performance Evaluation}
\end{table*}

\section{Experiment}

In this section, we describe the implementation of MOBIUS and the results of experiments using the MNIST dataset. 

\subsection{Implementation}
\label{implementation}

\begin{figure}[!t]
  \begin{center}
    \includegraphics[width=7cm]{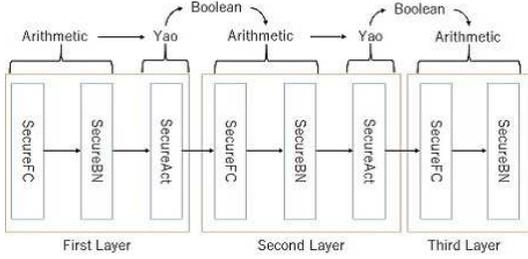}
  \caption{Flow of Implementation of MOBIUS}
  \label{fig:implement}
  \end{center}
\end{figure}

\begin{figure}[!t]
  \begin{center}
    \includegraphics[width=7cm]{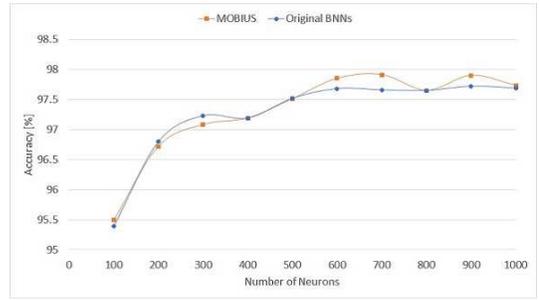}
  \caption{Prediction Accuracy of MOBIUS}
  \label{fig:result_accuracy}
  \end{center}
\end{figure}

\begin{figure}[!t]
  \begin{center}
    \includegraphics[width=7cm]{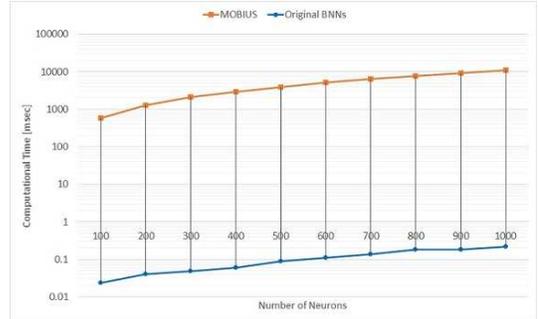}
  \caption{Computational Time of MOBIUS}
  \label{fig:result_comptime}
  \end{center}
\end{figure}

\subsubsection{Language and Library}
MOBIUS is implemented in C++ with the ABY library~\cite{ABY} for secure computation. The ABY library is a secure computation framework with two-party setting, and contains three types of shares, namely, \textit{Arithmetic}, \textit{Boolean}, and \textit{Yao}. 
These shares have different operations, and the ABY library provides efficient conversions between them. We refer the readers to the paper~\cite{ABY} for details on the ABY framework.

We briefly describe several parts related to our implementation below. 
The arithmetic shares can be used in arithmetic operations, such as addition and multiplication. 
Therefore, the secure full connection and secure batch normalization are implemented with arithmetic shares. 
In terms of share size, the ABY library includes four parameters as a modulo $p$, i.e., $8, 16, 32,$ and $64$ bits. 
Although we omit the detail due the space limitation, the MNIST dataset is available with $32$-bit parameter.
The secure activation requires a comparison operation of secure computation, which can be computed with Boolean or Yao shares. 
In the ABY library, Arithmetic shares cannot be directly converted into Boolean shares, i.e., the Arithmetic shares are first converted to Yao shares and then from Yao shares to Boolean shares. 
Therefore, since the conversion of Arithmetic shares to Boolean shares requires two conversions, we used Yao shares in the secure activation. Besides, according to the benchmark of the ABY library~\cite{ABY}, a comparison operation using Yao shares can be computed faster than using Boolean shares.
We hence implement the secure activation with Yao shares.

\subsubsection{Overview of Implementation}
We show the implementation flow of shares in Figure~\ref{fig:implement}. 
In the ABY library, Yao shares cannot be converted directly into Arithmetic shares, and thus Yao shares need to be converted to Boolean shares first, and then from Boolean shares to Arithmetic shares. We note that even though Boolean shares can be used in the secure activation, the flow shown in Figure~\ref{fig:implement} provides the fastest implementation.
In the algorithms of the original BNNs~\cite{BNN}, secure activation should be performed with Boolean shares because of the bit-shift operations. The extra conversion to Boolean shares and the overhead operations of Boolean shares may downgrade the computational performance.

The implementation of MOBIUS was created by simply using the available ABY library and is therefore not optimized unlike SecureML~\cite{SecureML}. 
Therefore, the performance of the following experiments can be improved by optimizing implementation. We plan to publish our source codes for subsequent works. 
The training phase is out of the scope of this work, and therefore a model is trained in advance. 
Shares of the model and input from a client are generated by the $\mathsf{PutSIMDINGate}$ function of the ABY library.

\subsection{Experimental Setting} \label{Data Sets}

\subsubsection{Machine Environments}
We conducted experiments with the MNIST dataset using the algorithms described in Section~\ref{BNN for MNIST} on two AmazonEC2 c4.8xlarge machines, both of which are running Linux and have 60 GB of RAM. 
The two machines are hosted in the same region as a LAN setting. The bandwidth is 1 GB/s, and the neural network has two hidden layers with 128 neurons in each layer. 
This setting is identical to that of SecureML~\cite{SecureML}. 
We also utilize the sign function as the activation function. The neural network is fully connected. 
We them compare the performance of our protocol with SecureML and other state-of-the-art protocols with cryptography~\cite{MiniONN,TAPAS,FHE-DiNN}.

\subsubsection{Dataset}

The MNIST dataset contains 70,000 images of handwritten digits from $0$ to $9$. 
In particular, the MNIST dataset has 60,000 training samples and 10,000 test samples, each with 784 features representing 28$\times$28 pixels in the image. 
Each feature is a grayscale between 0--255.

\subsection{Results} 
\label{result}
The experimental results are shown in Table~\ref{Performance Evaluation}, Figure~\ref{fig:result_accuracy}, and Figure~\ref{fig:result_comptime}. 
Table~\ref{Performance Evaluation} shows a comparison of different protocols based on capability to make a model oblivious, accuracy, and computational time. 
Figure~\ref{fig:result_accuracy} shows a comparison of MOBIUS and the original BNNs based on prediction accuracy with respect to the number of neurons. Figure~\ref{fig:result_comptime} shows a comparison of MOBIUS and the original BNNs based on computational time for prediction with respect to the number of neurons. 

As shown in Table~\ref{Performance Evaluation}, MOBIUS is the fastest system that combines BNNs and secure computation based on secret sharing despite having the capability to encrypt a model. 
We again note that the accuracy of MOBIUS may even be improved by optimizing our implementation. 
As shown in Figure~\ref{fig:result_accuracy}, MOBIUS has better prediction accuracy than the original BNNs because it uses improved BNNs that does not use bit-shift operations. 
Finally, as shown in Figure~\ref{fig:result_comptime}, the computational time of MOBIUS seems to be linear with respect to the number of neurons, although the computational time becomes 100 times longer than the original BNNs. We can thus approximately measure performance for any number of neurons.

\section{Conclusion}

In this work, we presented MOBIUS (Model-Oblivious BInarized neUral networkS), a system that enables scalable encrypted prediction and encryption of a trained model.
As our main technical contribution, we presented new algorithms of BNNs that are compatible with secure computation by representing all parameters in integers and removing the bit-shift method used in the original BNNs~\cite{BNN}. 
We then designed the main construction of MOBIUS with secure computation based on Arithmetic shares and Yao shares. 
We also conducted experiments using the MNIST dataset, and the results show that MOBIUS achieves higher computational performance and higher accuracy than SecureML, which is the only other system that considers the model-oblivious problem. 
As future work, we plan to conduct experiments on more complicated datasets, such as CIFAR10. 

\bibliographystyle{IEEEtran}
\bibliography{secure-bnn}
\end{document}